\documentclass[aps,showpacs,twocolumn,preprintnumbers,amsmath,amssymb,pra]{revtex4-1}

\usepackage{graphicx}
\usepackage{dcolumn}
\usepackage{bm}
\usepackage{color}

\newcommand{\Z}{\mathbb{Z}}

\newcommand{\rd}{\mathrm{d}}

\begin{document}

\title{Modulational instability in dispersion-kicked optical fibers}

\author{ S. Rota Nodari,$^1$ M. Conforti$^{2*}$, G. Dujardin$^{1,3}$, A. Kudlinski$^2$, A. Mussot$^2$, S. Trillo$^4$, and S. De Bi\`evre$^{1,3}$ }
\affiliation{$^1$Laboratoire Paul Painlev\'e, Universit\'e Lille 1, F-59655 Villeneuve d'Ascq, France,\\
$^2$PhLAM/IRCICA UMR 8523/USR 3380,  CNRS-Universit\'e Lille 1,  F-59655 Villeneuve d'Ascq, France,\\
$^3$\'Equipe MEPHYSTO, Inria Lille Nord-Europe Parc Scientifique de la Haute-Borne, F-59655 Villeneuve d’Ascq Cedex, France,\\
$^4$Dipartimento di Ingegneria, Universit\`a di Ferrara, Via Saragat 1, 44122 Ferrara, Italy.}
\email{matteo.conforti@univ-lille1.fr}

\begin{abstract}
We study, both theoretically and experimentally, modulational instability in optical fibers that have a longitudinal evolution of their dispersion in the form of a Dirac delta comb. By means of Floquet theory, we obtain an exact expression for the position of the gain bands, and we provide simple analytical estimates of the gain and of the bandwidths of those sidebands. An experimental validation of those results has been realized in several microstructured fibers specifically manufactured for that purpose. The dispersion landscape of those fibers is a comb of Gaussian pulses having widths much shorter than the period, which therefore approximate the ideal Dirac comb. Experimental spontaneous MI spectra recorded under quasi continuous wave excitation are in good agreement with the theory and with numerical simulations based on the generalized nonlinear Schr\"odinger equation.
\end{abstract}


\maketitle

\section{Introduction}
Modulational instability (MI) refers to a process where a weak periodic perturbation of an intense continuous wave (CW) grows exponentially as a result of the interplay between dispersion and nonlinearity. MI constitutes one of the most basic and widespread nonlinear phenomena in physics, and it has been studied extensively in several different physical systems like water waves, plasmas, and optical devices \cite{zakharov09}. For a cubic nonlinearity, as the one occurring in the nonlinear Schr\"odinger equation used to model optical fibers, the underlying physical mechanism can be understood in terms of four-wave mixing between the pump, signal and idler waves. However, the scalar four-wave interactions in a homogeneous fiber can be phase matched, and hence efficient, only in the anomalous group-velocity dispersion (GVD) regime. 
In the normal GVD regime, on the other hand, MI can occur in detuned cavities \cite{haelterman92, coen97}, thanks to constructive interference between the external driving and the recirculating pulse. Alternatively MI with normal GVD can also arise in systems with built-in periodic dispersion \cite{bronski96,smith96,Abdullaev96,abdullaev99}, among which dispersion oscillating fibers (DOFs) have recently attracted renewed attention \cite{armaroli12, droques12,droques13OL,droques13,finot13,conforti14}. 
In this case, phase matching relies on the additional momentum carried by the periodic dispersion grating (quasi-phase-matching). The occurrence of unstable frequency bands can then be explained using the theory of parametric resonance, a well-known instability phenomenon which occurs in linearized systems for which at least one parameter is varied periodically during the evolution \cite{armaroli12}. 
Up to now, most experimental investigations realised in optical fibers have been performed with basic sinusoidal ~\cite{droques12,droques13OL,droques13,finot13,conforti14} or amplitude modulated~\cite{copie2015} modulation formats. In this work, on the other hand, we study a radically different periodic modulation of the GVD, in the form of a periodic train (or comb) of Dirac delta spikes. This is a fundamental and widespread modulation format, encountered in  a variety of physical systems.  In optics, delta combs have been exploited to model lumped amplification in long haul fiber optic transmission systems  \cite{smith92,matera93},  or to model the power extraction in soliton based fiber lasers \cite{kelly92}. Moreover,  comb-like dispersion-profiled fibers have been exploited to generate trains of solitons starting from a beat signal \cite{taylor94}. At more fundamental level kicked systems are widely investigated as a paradigm for the emergence of chaos in perturbed Hamiltonian systems, with the delta-kicked rotor  being the most renowned example \cite{chirikov79}.
Its quantum version is described by a Schr\"odinger equation forced by a Dirac comb and has been extensively analyzed to study chaos in quantum systems \cite{casati79}.
 Recirculating fiber loops have been used to reproduce the quantum kicked rotor with an optical system, to study chaos and Anderson localization \cite{fischer99,atkins03}, and to illustrate how an optical system can be used to mimic other physical systems that are more difficult to reproduce experimentally. In the same vein, we hope the experimental setup we propose in this paper could be used as an experimental platform to investigate such phenomena in the presence of nonlinearities, a topic of much current interest. Finally the approach that we propose to analyse MI in the fiber with delta-kicked GVD allows us, on one hand to enlighten the featuress of the parametric resonance that are not dependent on the specific format of the modulation, and on the other hand to compare and contrast the features of the ideal delta-kicked profile with other formats including non-ideal (physically realizable) kicking as well as widely employed profiles such as oscillating GVD.

The paper is organized as follows. In Section~\ref{s:centralfreq} we provide a simple argument allowing to determine the central frequencies of the unstable sidebands for general periodically modulated fibers. In Section~\ref{s:diraccomb}, we then use Floquet theory to analytically compute the width of the gain bands and as well as their maximum gain for dispersion-kicked fibers. In Section~\ref{s:approxdelta} we investigate numerically the effect of the smoothing of the delta comb. In Section~\ref{s:expresults}, we describe the experimental set-up and we compare the experimental results with theory and numerical simulations based on the generalized nonlinear Schr\"odinger equation. We draw our conclusions in Section~\ref{s:conclusions}.

\section{Identifying the gain band central frequencies}\label{s:centralfreq}
Consider the NLSE 
\begin{equation}
	\label{nlsefiberrot}
	i\frac{\partial u}{\partial z}-\frac{\beta_2(z)}{2}\frac{\partial^2u}{\partial t^2}+\gamma(z)|u|^2u=0.
\end{equation}
We will assume the dispersion $\beta_2(z)$ and the nonlinearity coefficient $\gamma(z)$ are of the form
\begin{equation}\label{eq:beta2}
\beta_2(z)=\beta_{\mathrm{av}}+\beta_m f_{Z}(z),\quad \gamma(z)=\gamma_{\mathrm{av}}+\gamma_m g_Z(z),
\end{equation}
where $f_{Z}$ and $g_Z$ are periodic functions of period $Z$ such that $\min f_Z=-1=\min g_Z$, and $\int_{-Z/2}^{Z/2} f_Z(z)dz=\int_{-Z/2}^{Z/2} g_Z(z)dz=0$. 

\begin{figure}
\begin{center}
\includegraphics[width=8cm]{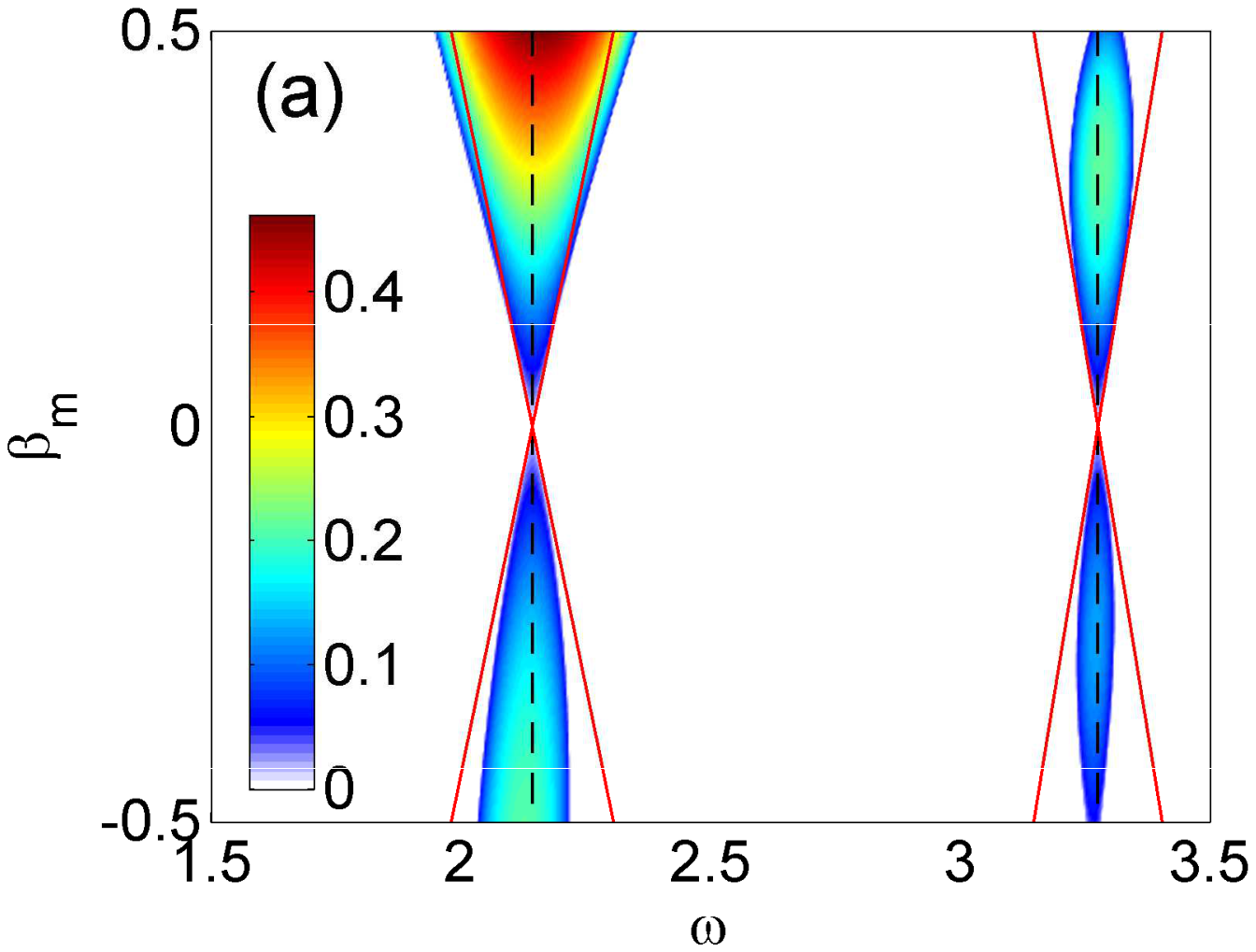}
\includegraphics[width=8cm]{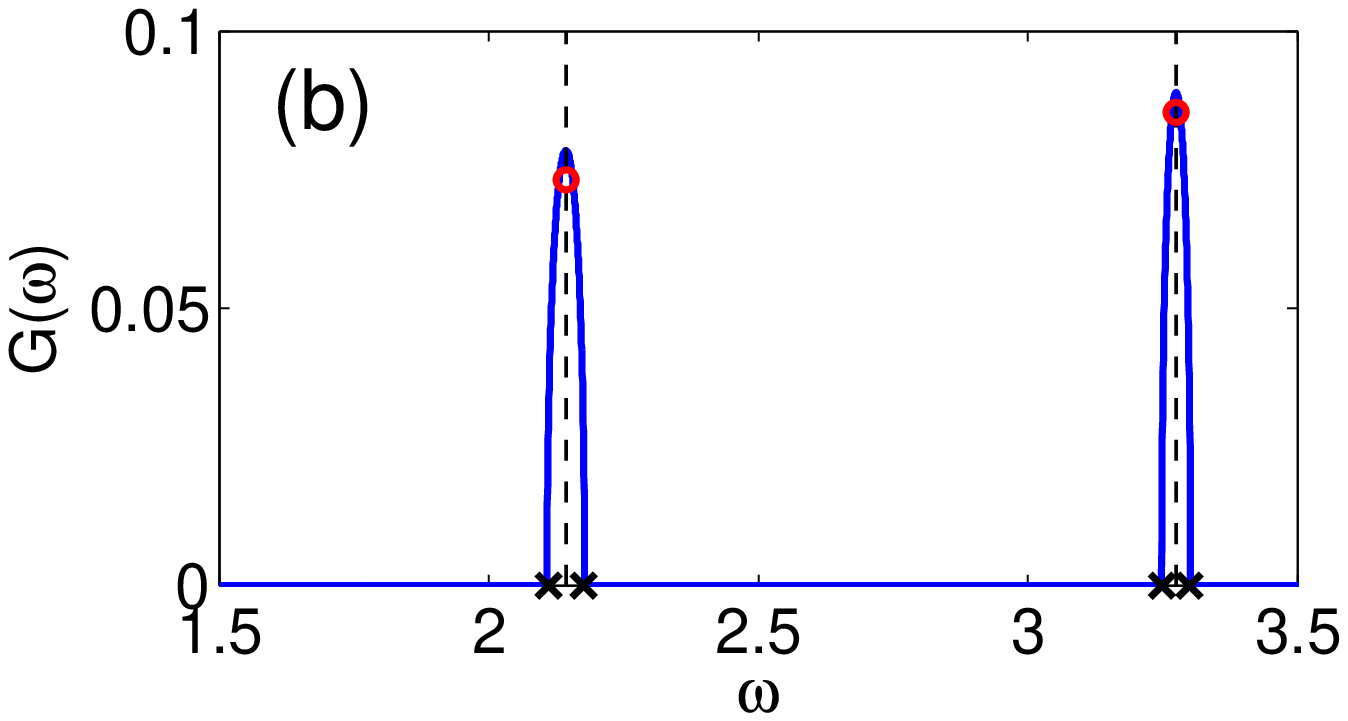}
\includegraphics[width=8cm]{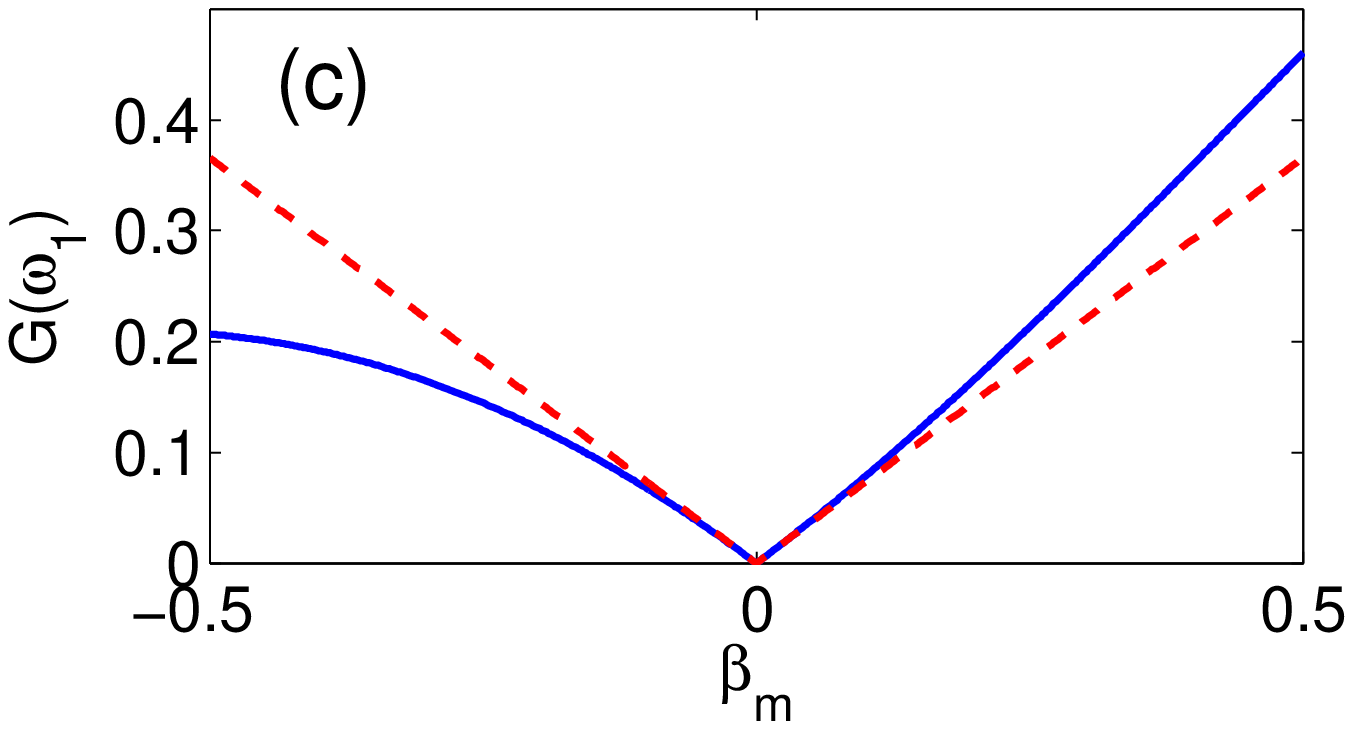}
\caption{ (a) Level plot of the gain $G(\omega, \beta_m)=\ln\left(\max\{|\lambda_+|, |\lambda_-|\}\right)/Z$ in the $(\omega, \beta_m)$ plane, for $\beta_{\mathrm{av}}=1$, $Z=1$, $\gamma_{\mathrm{av}}=1$ and $P=1$.  The dashed black lines corresponds to the tips of the Arnold tongues (\ref{omega_l}) at $\omega=\omega_1=2.1433$ and $\omega_2=3.2748 $.  The solid red lines corresponds to the gain bandwidth, which can be computed from~\eqref{band}. (b) MI gain for $\beta_m=0.1$; Red circles, estimates of maximum gain (\ref{gmax}); Black crosses, estimates of the bandwidth \eqref{band}. (c) Solid blue curve, MI gain for $\omega=\omega_1$; Dashed red curve, approximation of maximum gain (\ref{gmax}). \label{fig:arnoldtongues}}
\end{center}
\end{figure}

Let $u_0(z)=\sqrt P\exp(iP\int_0^z\gamma(z')\rd z')$ be a stationary solution of~\eqref{nlsefiberrot}. We consider a perturbation of $u_0(z)$ in the form $u(z,t)=(v(z,t)+1)u_0(z)$,  where the perturbation $v(z,t)$ satisfies $|v|\ll 1$. Inserting this expression in~\eqref{nlsefiberrot}, and retaining only the linear terms we find
\begin{equation}
	\label{nlsefiberlin}
	i\frac{\partial v}{\partial z}-\frac{\beta_2(z)}{2}\frac{\partial^2 v}{\partial t^2}+\gamma(z) P(v+v^*)=0.
\end{equation}
Writing $v=q+ip$, with $q$ and $p$ real functions, we obtain the following linear system:
\begin{equation*}
	\left\{
	\begin{aligned}
		&\frac{\partial q}{\partial z}-\frac{\beta_2(z)}{2}\frac{\partial^2p}{\partial t^2}=0,\\
		&\frac{\partial p}{\partial z}+\frac{\beta_2(z)}{2}\frac{\partial^2q}{\partial t^2}-2\gamma(z) P q=0.
	\end{aligned}
	\right.
\end{equation*}
Finally, taking the Fourier transform of this system in the time variable $t$, leads to 
\begin{equation}
	\label{nlsefiberlinfouriersys}
	\left\{
	\begin{aligned}
		&\frac{\partial \hat q }{\partial z}+\frac{\beta_2(z)}{2}\omega^2\hat p=0,\\
		&\frac{\partial \hat p}{\partial z}-\frac{\beta_2(z)}{2}\omega^2\hat q-2\gamma(z) P\hat q=0,
	\end{aligned}
	\right.
\end{equation}
where we used the definiton $\hat q(z,\omega)=\frac{1}{\sqrt{2\pi}}\int q(z,t)e^{-i\omega t}\,\rd t$. Note that this is a Hamiltonian dynamical system in a two-dimensional phase plane with canonical coordinates $(\hat q, \hat p)$. 
Analyzing the linear (in)stability of the stationary solution $u_0(z)$ therefore reduces to studying the solutions to~\eqref{nlsefiberlinfouriersys} for each $\omega$. Since the coefficients in the equation are $z$-periodic with period $Z$, Floquet theory applies. This amounts to studying the linearized evolution over one period $Z$, to obtain the Floquet map $\Phi_{\beta_m, \gamma_m}$ which in the present situation is the two by two real matrix defined by $\Phi_{\beta_m, \gamma_m}^{\mathrm{lin}}(\hat q(0), \hat p(0))=(\hat q(Z), \hat p(Z))$. As a result $(\hat q(nZ), \hat p(nZ))={\Phi_{\beta_m, \gamma_m}^{\mathrm{lin}}}^n(\hat q(0), \hat p(0))$. Note that $\Phi_{\beta_m, \gamma_m}^{\mathrm{lin}}$  necessarily has determinant one, since it is obtained by integrating a Hamiltonian dynamics, of which we know that it preserves phase space volume. As a consequence, if $\lambda$ is one of its eigenvalues, then so are both its complex conjugate $\lambda^*$ and its inverse $\lambda^{-1}$. This constrains the two eigenvalues of $\Phi_{\beta_m, \gamma_m}^{\mathrm lin}$ considerably: they are either both real, or lie both on the unit circle. Now, the dynamics is unstable only if  there is one eigenvalue $\lambda$ satisfying $|\lambda|>1$, in which case both eigenvalues are real. We will write $\lambda_\pm$ for the two eigenvalues of $\Phi_{\beta_m, \gamma_m}^{\mathrm{lin}}$. We are interested in studying the gain, that is
\begin{equation}\label{gdef}
G(\omega, \beta_m, \gamma_m)=\ln\left(\max\{|\lambda_+|, |\lambda_-|\}\right)/Z
\end{equation}
as a function of $\omega$, $\beta_m$ and $\gamma_m$. It measures the growth of $(\hat q(nZ), \hat p(nZ))$.  The gain vanishes if the two eigenvalues lie on the unit circle. A contour plot of the gain in the $(\omega, \beta_{m})$ plane, for the case of the delta comb dispersion modulation that is the main subject of this paper, can be found in Fig.~\ref{fig:arnoldtongues}. The regions where the gain does not vanish are commonly referred to as Arnold tongues. We will explain below that, whereas their precise form depends on the choice of $f_Z, g_Z$, the position of their tips does not.

Since the system~\eqref{nlsefiberlinfouriersys} is not autonomous, it cannot be solved analytically in general. Nevertheless, the above observations will allow us to obtain some information about its (in)stability for small $\beta_m$, $\gamma_m$, and valid for all perturbations $f_Z, g_Z$, whatever their specific form. 

To see this, we first consider the case $\beta_m=0=\gamma_m$. It is then straightforward to integrate the system \eqref{nlsefiberlinfouriersys}. The linearized Floquet map is then given by
\begin{equation}\label{floq_lin}
\Phi_{Z, 0}^{\mathrm{lin}}=
\begin{pmatrix}
\cos(kZ)&
-\frac{\frac{\overline{\beta}_2}{2}\omega^2}{k}\sin(kZ)\\
\frac{k}{\frac{\overline{\beta}_2}{2}\omega^2}\sin(kZ)&\cos(kZ)
\end{pmatrix}:=L,
\end{equation}
where
\begin{equation}\label{eq:Theta}
k^2=\frac{\overline\beta_2}{2}\omega^2\left(\frac{\overline\beta_2}{2}\omega^2+2\gamma_{\mathrm av} P\right).
\end{equation}
Here $\overline\beta_2=\beta_{\mathrm{av}}>0$  (normal average dispersion), since we restrict our investigations to the defocusing NLS. Note that the matrix $L$ has determinant equal to $1$, as expected. The eigenvalues of $L$ can be readily computed as
\begin{equation}\label{eq:lambdapm0}
\lambda_{\pm}(\omega, \beta_m=0=\gamma_m)=\exp(\pm ikZ).
\end{equation}
%
What will happen if we now switch on the interaction terms $f_Z(z)$ and $g_Z(z)$? It is then no longer possible, in general, to give a simple closed form expression of the solution to~\eqref{nlsefiberlinfouriersys}, which is no longer autonomous, and hence of the linearized Floquet map $\Phi_{Z, \beta_m, \gamma_m}^{\mathrm{lin}}$.    Nevertheless, we do know that, for small $\beta_m, \gamma_m$, the eigenvalues of  $\Phi_{Z, \beta_m, \gamma_m}^{\mathrm{lin}}$  must be close to the eigenvalues $\lambda_\pm(\omega, \beta_m=0=\gamma_m)$. We then have two cases to consider.

\begin{figure}
\begin{center}
\includegraphics[width=9cm]{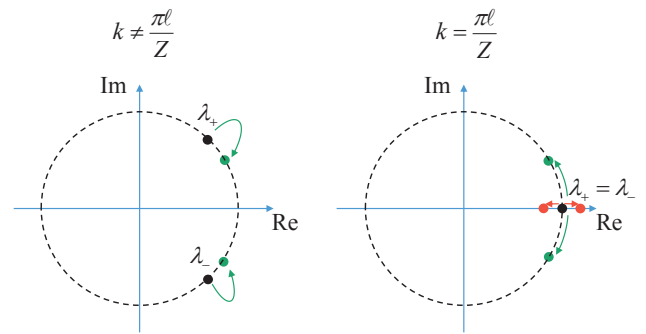}
\caption{Sketch illustrating, in the complex plane, the effect of the interaction terms $f_Z(z)$ and $g_Z(z)$ on the eigenvalues of the linearized Floquet map (\ref{floq_lin}). Black dots correspond to the unperturbed eigenvalues lying on the unit circle (dashed line). Coloured dots show the new position of the eigenvalues after switching on the perturbations, leading to a stable regime when $k\not=\frac{\pi\ell}{Z}$ and an unstable one when $k=\frac{\pi\ell}{Z}.$ }\label{eigenvalues}
\end{center}
\end{figure}

\noindent{\bf Case 1. $k\not=\frac{\pi \ell}{Z}, \ell\in\Z$}.
Now $\lambda_-(\omega, \beta_m=0=\gamma_m)=\lambda^*_+(\omega, \beta_m=0=\gamma_m)$, \emph{they are distinct}, and they both lie on the unit circle, away from the real axis. They then must remain on the unit circle under perturbation since, for the reasons explained above, they cannot move into the complex plane away from the unit circle. 
Consequently, in this case, the stationary solution $u_0(z)$ is linearly stable under a sufficiently small perturbation by $\beta_mf_Z(z)$ and $\gamma_m g_Z(z)$, and this statement does not depend on the precise form of $f_Z(z)$ or of $g_Z(z)$. In fact, with growing $\beta_m$ and/or $\gamma_m$, the two eigenvalues will move along the unit circle until they meet either at $-1$ or at $+1$ for some critical value of the perturbation parameters. Only for values of the latter above that critical value can the system become unstable. A pictorial description of this situation is shown in the left hand side of Fig.~\ref{eigenvalues}. \\

\noindent{\bf Case 2. $k=\frac{\pi \ell}{Z}$, $\ell\in\Z$.}
Now $\lambda_{+}=\lambda_{-}=\pm 1$ is a doubly degenerate eigenvalue of $\Phi_{Z, 0}^{\mathrm{lin}}$. Under a small perturbation, the degeneracy can be lifted and two real eigenvalues can be created, one greater than one, one less than one in absolute value. The system has then become unstable! Of course, it will now depend on the type of perturbation whether the system becomes unstable, remains marginally stable (the two eigenvalues don't move at all, but stay at $1$ or $-1$), or  becomes stable (the two eigenvalues move in opposite directions along the unit circle).   A pictorial description of this situation is shown in the right hand side of Fig.~\ref{eigenvalues}. For the Dirac comb modulation of $\beta_2(z)$, which is our main object of study in this paper, the details are given in the next section. 

In conclusion, examining~\eqref{eq:Theta}, one sees that only if $\omega=\omega_\ell$, where
\begin{equation}\label{omega_l}
\omega_\ell^2=\frac{2}{\beta_\mathrm{av}}\left(\sqrt{(\gamma_{\mathrm{av}} P)^2+\left(\frac{\ell\pi}{Z}\right)^2}-\gamma_{\mathrm av} P \right),
\end{equation}
can an infinitely small Hamiltonian perturbation of $\Phi_{Z, 0}^{\mathrm{lin}}$ lead to an unstable linearized dynamics near the fixed points $u_0(z)$ considered.  
These values of $\omega$ therefore correspond to the tips of the Arnold tongues, that is, to the positions of the (centers of) the unstable sidebands of the defocusing NLS under a general periodic perturbation $f_Z, g_Z$. This is illustrated for a Dirac comb modulation of the GVD in Fig.~\ref{fig:arnoldtongues}. One also observes in that figure that, for a value of $\omega$ close to some $\omega_\ell$, the system becomes unstable only for a small but nonzero critical value of $\beta_m$, that we shall compute below for the Dirac delta comb GVD.

Equation (\ref{omega_l}) was derived in~\cite{armaroli12} by appealing to the theory of parametric resonance and Poincar\'e-Lindstedt perturbation theory. Our argument above is elementary and shows  in a simple manner that the resonant frequencies $\omega_\ell$ do not at all depend on the form of $f_Z$ or $g_Z$. Note that, if $f_Z(z)=\sin(\frac{2\pi}{Z}z)$ and $g_Z(z)=0$, a case considered in~\cite{droques13OL}-\cite{droques13}, the system~\eqref{nlsefiberlinfouriersys} is equivalent to the equation of a harmonic oscillator of (spatial) frequency $k$, sinusoidally modulated with period $Z$. In that case the system leads to a Mathieu equation for which it is known that resonance occurs when the period of the modulation is a integer multiple of the half (spatial) period of the oscillator, which is $2\pi/k$.

Additional physical insight can be obtained by expanding Equation~(\ref{omega_l}) for small power, i.e. assuming $\gamma_{av}P\ll |\ell|\pi/Z$. At zero order we recover the well known quasi-phase-matching relation \cite{smith96,droques12,matera93}
\begin{equation}\label{phasematching}
\beta_\mathrm{av}\omega_\ell^2+2\gamma_{av}P=\frac{2 \pi \ell}{Z}.
\end{equation}
Equation (\ref{phasematching}) entails the  conservation of the momentum, made possible thanks to the virtual momentum carried by the dispersion grating, of the four wave mixing interaction between two photons from the pump, going into two photons in the symmetric unstable bands at lower (Stokes) and higher (antiStokes) frequencies with respect to the pump.

\section{Calculation of the Modulational Instability gain bands: Dirac comb}\label{s:diraccomb}
We now turn our attention to the computation of the gain $G(\omega)$, in particular for values of $\omega$ close to the resonant frequencies. We concentrate on the special case where
the GVD is a Dirac delta comb:
\begin{equation}\label{eq:diracformfLambda}
f_Z(z)=\left[\sum_{n\in\Z} \delta (z/Z-n)\right]-1,\qquad \gamma_m=0.
\end{equation}
Since in the rest of this paper, $\gamma_m=0$, we will drop it from the notation. To compute the gain, we need to compute the linearized dynamics $\Phi_{\beta_m}^{\mathrm{lin}}$ and determine the behaviour of its eigenvalues $\lambda_\pm( \beta_m, \omega)$ in the neighbourhood of $\beta_m=0$ and $\omega=\omega_\ell$ in the $(\omega, \beta_m)$-plane. 

In this case the linearized Floquet map is easily seen to be explicitly given by
\begin{equation}\label{eq:linfloqbetam}
\Phi_{\beta_m}^{\mathrm{lin}}= 
LK
\end{equation}
where
$L$ is defined by Equation~(\ref{floq_lin}), but now with $\overline\beta_2=\left[\beta_{\mathrm{av}}-\beta_m\right]$, and 
\begin{equation}
K=
\begin{pmatrix}
\cos\left(\beta_m\frac{\omega^2}{2}Z\right)&-\sin\left(\beta_m\frac{\omega^2}{2}Z\right)\\
\sin\left(\beta_m\frac{\omega^2}{2}Z\right)&\cos\left(\beta_m\frac{\omega^2}{2}Z\right)
\end{pmatrix}.
\end{equation}
 The characteristic polynomial of $LK$ is given by 
\begin{equation*}
\lambda^2-2\delta(\omega,\beta_m)\lambda+1=0,
\end{equation*}
so that the eigenvalues of (\ref{eq:linfloqbetam}) can be computed explicitly as:
\begin{equation}\label{eq:floqeigenvalues}
\lambda_{\pm}(\omega,\beta_m)=\delta(\omega,\beta_m)\pm\sqrt{\delta(\omega,\beta_m)^2-1},
\end{equation}
with
$$
\delta=\cos(kZ)\cos\left(\theta\right)-\frac{\frac{\overline \beta_2}{2}\omega^2+\gamma_{av} P}{k}\sin(kZ)\sin\left(\theta\right),
$$
and $\theta=\beta_m\frac{\omega^2}{2}Z$.

A Taylor expansion of $\delta(\omega,\beta_m)$ about $(\omega_\ell, 0)$ yields
\begin{equation}\label{eq:arnoldtongueslin}
\delta(\beta_m,\Omega)\simeq (-1)^\ell\left[1+ C_\ell \beta_m^2 - D_\ell (\omega-\omega_\ell)^2\right],
\end{equation}
where
\begin{align}
C_\ell&= Z^2\left(\frac{\omega_\ell^2}{2}\right)^2\left(\frac{Z}{\pi\ell}\right)^2(\gamma_{av} P)^2,\\
D_\ell&=\beta_\mathrm{av}^2Z^2\omega_\ell^2\left(\left(\frac{Z}{\pi\ell}\right)^2(\gamma_{av} P)^2+1\right).
\end{align}
The dependence in $\beta_m^2$ (not in $\beta_m$) entails that the sign of the kick has no incidence in this regime, i.e. assuming $|\beta_m|\ll 1$.

 Formula (\ref{eq:arnoldtongueslin}) shows that $(0,\omega_\ell)$ is a saddle point for $\delta(\beta_m, \omega)$. If $\ell$ is even, $\lambda_+(\beta_m,\omega)>1$ occurs close to $(0,\omega_\ell)$, and if $\ell$ is odd, $\lambda_-(\beta_m,\omega)<-1$  close to $(0,\omega_\ell)$. More precisely, 
\begin{equation*}
	\max\left(|\lambda_+|,|\lambda_-|\right)=1+\sqrt{C_\ell \beta_m^2 - D_\ell (\omega-\omega_\ell)^2}
\end{equation*}
from which we can find an estimate of the gain amplitude $G(\omega_\ell, \beta_m)$ and  of the bandwidth $B(\omega_\ell, \beta_m)$ near the tips of the tongue at $\omega_\ell$, as:
\begin{equation}\label{gmax}
G(\omega_\ell, \beta_m)\approx|\beta_{m}|\frac{\omega^2_\ell}{2}\frac{Z}{\pi\ell}\gamma_{av} P,
\end{equation}
\begin{equation}\label{band}
B(\omega_\ell,\beta_m)=\frac{|\beta_{m}|}{\beta_{\mathrm{av}}}\frac{\omega_\ell}{2}\frac{\gamma_{av} P}{\sqrt{\left(\frac{\pi \ell}{Z}\right)^2+(\gamma_{av} P)^2}}
\end{equation}
Note that the threshold value for $\beta_m$ above which instability occurs can be read off from the above by  setting $|\delta(\beta_m,\omega)|=1$ which corresponds to 
$$
|\beta_m|\geq \sqrt{\frac{D_\ell}{C_\ell}}|\omega-\omega_\ell|.
$$ 
This confirms again, as expected, that an arbitrary small $\beta_m$ will generate instability right at $\omega=\omega_\ell$. 
In Fig. \ref{fig:arnoldtongues}a we show an example of the analytically computed MI gain, showing the first two Arnold tongues. As can be seen, for a small enough strength of perturbation, let's say $|\beta_m|\le0.1$, the approximation (\ref{band}) gives a good estimate of the width of the parametric resonance (see red curves). This situation is detailed further in Figs. \ref{fig:arnoldtongues}b,c, showing a section for $\beta_m=0.1$ and $\omega=\omega_1$, respectively.

Finally, a straightforward calculation gives the asymptotic behaviour of the gain $G$ at $\omega_\ell$ for $\ell$ large and $\beta_m$ fixed, that is
\begin{equation}\label{gasym}
G(\omega_{\ell}, \beta_m)\approx 
\frac{\sqrt{4\beta_{\mathrm{av}}^2\sin^2(\alpha(\ell))(\gamma_{\mathrm{av}}P)^2-\beta_m^2(\gamma_{\mathrm{av}}P)^4Z^2}}{2|\overline\beta_2|\pi\ell}
\end{equation}
with $\alpha(\ell)=\frac{\beta_m}{\beta_{\mathrm{av}}}(\pi\ell-\gamma_{\mathrm{av}}PZ)$ whenever $4\beta_{\mathrm{av}}^2\sin^2(\alpha(\ell))-\beta_m^2(\gamma_{\mathrm{av}}P)^2Z^2>0$, and $G(\omega_\ell)\approx 0$ otherwise.

\begin{figure}
\begin{center}
\includegraphics[width=8cm]{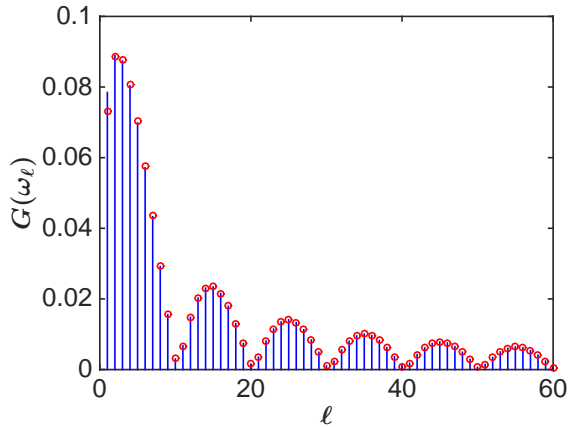}
\caption{ MI gain for $\omega=\omega_\ell$ as a function of $\ell$; Red circles, estimated gain given by \eqref{gasym}. Parameter values are $\beta_{\mathrm{av}}=1$, $Z=1$, $\gamma_{\mathrm{av}}=1$, $P=1$,  $\beta_m=0.1$. }\label{profile_gain_large}
\end{center}
\end{figure}

In Fig. \ref{profile_gain_large}, we show an example of the analytically computed MI gain at $\omega_\ell$ as a function of $\ell$. We compare it to the approximation \eqref{gasym}, which is very accurate, even for small $\ell$ (see red circles). Note in particular that the oscillating behaviour of the gain is well captured by \eqref{gasym} which, for $\ell$ large enough and $\beta_m$ small, can be approximated by  
\begin{equation*}
G(\omega_\ell,\beta_m)\simeq|\beta_m|\left|\frac{\sin\left(\frac{\beta_m}{\beta_{\mathrm{av}}}\pi\ell-\frac{\beta_m}{\beta_{\mathrm{av}}}\gamma_{\mathrm{av}}PZ\right)}{\frac{\beta_m}{\beta_{\mathrm{av}}}\pi\ell}\right|.
\end{equation*}

\noindent{\bf Summing up.} It is clear from the above that, precisely at the values $\omega_\ell$, which only depend on $Z$ and on $\gamma_{av} P$, but not on the precise form of $f_Z$, any small perturbation can create an instability and hence a gain. At frequencies $\omega$ near these particular values, a minimal threshold strength of $\beta_m$ is needed to create an instability. This minimal value, and even the fact that an instability is indeed generated, does depend on the precise form of $f_Z$. For the Dirac comb the explicit expression for the gain in this regime can be read off from~\eqref{gasym}. 
\begin{figure}
\begin{center}
\includegraphics[width=8cm]{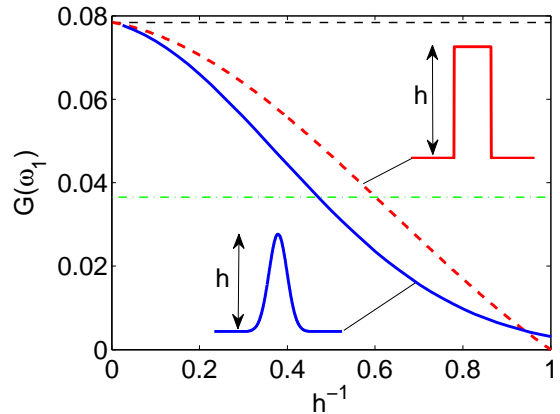}
\caption{ Gain $G(\omega_1):= G(\omega_1, \beta_m)$, with $\beta_m=0.1$, calculated numerically at the first parametric resonant frequency  $\omega_1$, for two approximations of delta functions as a function of their inverse height $h^{-1}$. Solid blue curve, gaussian approximation; Dashed red curve, rectangular pulse approximation; Dashed horizontal line, Dirac delta limit; Dash-dotted horizontal line, sinusoidal modulation. Parameter values are $\beta_\mathrm{av}=1$, $Z=1$, $\gamma=1$, $P=1$. }\label{delta_non_perf}
\end{center}
\end{figure}


\section{Approximations of the delta function}\label{s:approxdelta}
In order to shed light on the dependence of the gain on the shape of the periodic modulation, and also with an eye towards the experimental realization of the Dirac comb fiber,   we now analyze what happens when the Dirac comb is approximated by a train of physically realizable ``kicks''.
We thus consider a smoothened Dirac comb described by
\begin{equation}
f_Z(z)=\left(\sum_{n\in\Z} \eta(z-nZ)\right)-1,
\end{equation}
where we normalize the positive function $\eta(z)$ in order to have $\int_{-Z/2}^{Z/2} f_Z(z)dz=0$. 
For a rectangular pulse of width $w<Z$, we get
\begin{equation}
\eta(z)=\left\{
\begin{array}{cc}
h=\frac{Z}{w},~&-\frac{w}{2} < z <\frac{w}{2} \\
0,~& \mathrm{elsewhere}. 
\end{array}
\right.
\end{equation}

For a gaussian function $\eta(z)=h\exp(-z^2/(2w^2))$, the maximum amplitude of the kick can be calculated as 
$$h=\sum_{n\in\Z}\frac{Z}{w \sqrt{\frac{\pi}{2}} \left(\mathrm{Erf}\left(\frac{(1 - 2 n) Z}{2 \sqrt{2} w}\right) +   \mathrm{Erf}\left(\frac{(1 + 2 n) Z}{2 \sqrt{2} w}\right) \right)},$$
that in the limit $w\ll Z$ gives
$$h\approx\frac{Z}{w\sqrt{2\pi}}.$$
Note that, in these models, we have ($\beta_m\geq0$)
$$
\beta_{\mathrm{av}}-\beta_m\leq \beta_{\mathrm{min}}\leq \beta_{\mathrm{av}},\quad \beta_{\mathrm{max}}=\beta_{\mathrm{av}}+(h-1)\beta_m.
$$
Hence $h=2$ corresponds to a rather symmetric situation where $\beta_{\mathrm{av}}$ is close to the midpoint between $\beta_{\mathrm{min}}$ and $\beta_{\mathrm{max}}$, so that $\beta_2(z)$ fluctuates symmetrically about its average value. Whereas $h>>1$ corresponds to a very asymmetric situation where $\beta_2(z)$ has a large abrupt peak. The parameter $h$ therefore controls the shape of the GVD modulation at fixed $\beta_{\mathrm{av}}$ and $\beta_{\min}$ (or $\beta_{m}$).

As shown in the previous section, by changing the shape of the kick, we do not change the frequency of the parametric resonances. The smoothing of the delta function nevertheless does modify the characteristics of the MI by changing the value of the gain, as we now illustrate by computing the gain numerically at the resonant frequencies $\omega_\ell$. An example of how the changing shape of the modulation $f_Z$ modifies the first parametric resonance is illustrated in Fig. \ref{delta_non_perf}, that shows the gain $G(\omega_1, \beta_m)$ at fixed $\omega_1$ and $\beta_m$, as a function of the peak amplitude $h$ (or, equivalently, the width $w$) of the kicks. We make the following observations. First, a good approximation of the gain given by the Dirac comb is obtained for $h>10$, both for the rectangular and gaussian pulses. Second, for $h=1$, the gain of the square pulse modulation is zero, as expected, since we  are then in the limit case of a constant modulation (and normal GVD). Third, it is apparent that the Dirac comb gives the highest possible gain, for a fixed area of the kicks and fixed $\beta_m$ and $\beta_{\mathrm{av}}$. Finally, it is interesting to note that a sinusoidal modulation $f_Z(z)=\sin(\frac{2\pi}{Z}z)$, with the same value of $\beta_m$, gives a gain close to one half with respect to the delta case.
Indeed for a sinusoidal modulation, it has been shown that (see Equation~(7) from Ref. \cite{droques13})
\begin{equation}\label{gmax_sin}
G_{\sin}(\omega_\ell, \beta_m)=\left|J_\ell\left(\beta_{m}\frac{\omega^2_\ell}{2}\frac{Z}{\pi\ell}\right)\right|\gamma_{\mathrm{av}} P.
\end{equation}
By expanding Equation~(\ref{gmax_sin}) for small $\beta_m$, we get
\begin{equation}
G_{\sin}(\omega_\ell, \beta_m) = \frac{1}{2}G(\omega_\ell, \beta_m),
\end{equation}
at first order in $\beta_m$. In conclusion, a large concentrated perturbation of the GVD about its average enhances the MI gain.

\section{Experimental results}\label{s:expresults}

\begin{figure}
\begin{center}
\includegraphics[width=8cm]{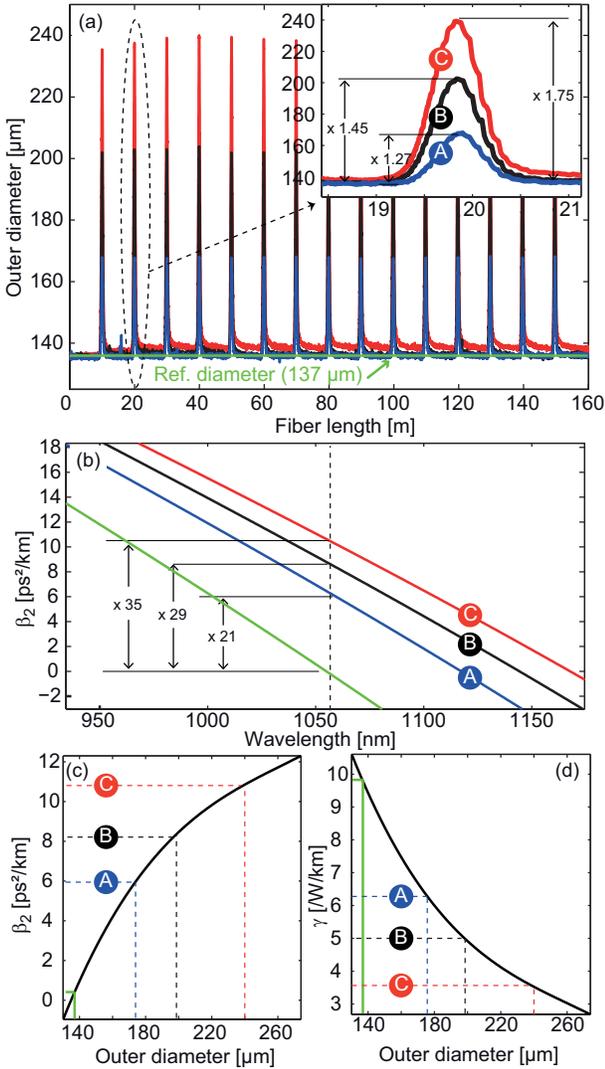}
\caption{(a) Outer diameter evolution versus fiber length for fibers A, B and C. Inset: Snapshot on one leitmotiv. (b) Calculated GVD curves corresponding to the background diameter (green curve) and the maximum diameter values of fibers A (blue curve), B (black curve) and C (red curve). (c) GVD and nonlinear coefficients evolution versus the diameter at 1052.5 nm. Curves in (b), (c) and in (d) calculated from Ref.\cite{saitho} with $d/\Lambda=0.48$ and $\Lambda_{min}=3.445 \mu m$. \label{fig:fibres exp}}
\end{center}
\end{figure}

\begin{figure}
\begin{center}
\includegraphics[width=8cm]{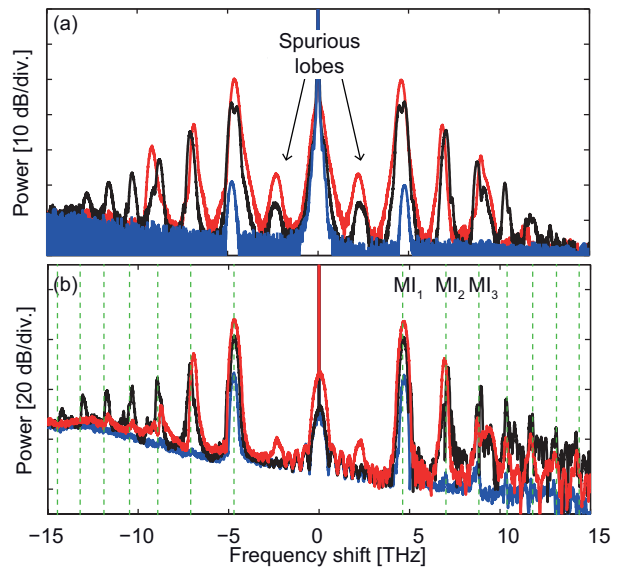}
\caption{Spectra recorded at the output of fibers A, B and C. We used the same color label than in Fig. \ref{fig:fibres exp}.(a) Experiments and (b) numerical simulations. Parameters: $\beta_3=5.6\times 10^{-41}s^3/m, \beta_4=-1\times 10^{-55}s^4/m, \alpha=5 dB/km$. \label{fig:exp et simul}}
\end{center}
\end{figure}

It is well known that in homogeneous fibers, the GVD depends on the diameter of the fiber. One can therefore modulate the GVD by modulating the diameter of the fibers as a function of $z$, as in~\cite{droques12, droques13OL, droques13}. We manufactured three different microstructured optical fibers modulated by a series of Gaussian pulses to approximate the ideal Dirac delta comb studied in Section~\ref{s:diraccomb}. The change of their outer diameters $d(z)$ along the fiber is represented in Fig. \ref{fig:fibres exp}(a). As can be seen in the inset, their diameters have a gaussian shape with a standard deviation $w$, which is the same for all three fibers, and very small ($w\simeq 0.14$ m) 
compared to the period of the comb (10 m). Hence we can write
$$
d(z)=d_{\min}+\delta g_w(z)\leq d_{\max}=d_{\min}+\delta,
$$
where $g_w(z)=\sum_n\exp(-\frac12\left(\frac{z-nZ}{w}\right)^2)$. The three fibers have a very similar minimum diameter $d_{\min}=137 {\mu}m$, while their maximum values are different. We have $d_{\max}^A\simeq$172 $\mu m$, $d_{\max}^B\simeq$207 $\mu m$ and $d_{\max}^C\simeq$240 $\mu m$ for fibers labelled A, B and C, respectively, corresponding to $\delta^A=35\mu m, \delta^B= 70 \mu m, \delta^C=103 \mu m$. To understand how the two experimental parameters $w$ and $\delta$ control the quality of the approximation of the delta function on the one hand,  and the value of $\beta_m$ on the other hand, we proceed as follows. First, $d_{\mathrm{av}}=d_0+\frac{\delta}{Z}w\sqrt{2\pi}$, so that $d_{\mathrm{av}}^A=139.1, d_{\mathrm{av}}^B=141.2, d_{\mathrm{av}}^C=143.2 $. A first order Taylor expansion of $\beta_2(d)$ about $d_{\mathrm{av}}$ yields
$$
\beta_2(z)=\beta_2(d(z))\simeq \beta_2(d_{\mathrm{av}}) +\beta_2'(d_{\mathrm{av}})\delta\left[g_w(z)-\frac{w}{Z}\sqrt{2\pi}\right].
$$
Comparing this to~\eqref{eq:beta2} we find
$$
\beta_{\mathrm{av}}=\beta_2(d_{\mathrm{av}}),\ \beta_m=\beta_2'(d_{\mathrm{av}})\frac{\delta w}{Z}\sqrt{2\pi}, 
$$
and
$$
\ f_Z(z)=\frac{Z}{w\sqrt{2\pi}}g_w(z)-1.
$$
Hence, with the notation of Section~\ref{s:approxdelta}, $h=\frac{Z}{w\sqrt{2\pi}}\simeq 28.5$. This corresponds to  $h^{-1}\approx 0.03$  proving that these Gaussian pulses should induce a very similar parametric gain compared to ideal Dirac delta functions (see Fig. \ref{delta_non_perf}). 
Furthermore, the height $\delta$ of the Gaussian pulse controls $\beta_m$, which will allow us to investigate the impact of this parameter on the first MI side lobe gain, as it was done in the theoretical study and illustrated in Fig.\ref{fig:arnoldtongues}. 
For all three fibers, the ratio of the diameter of the holes over $\Lambda$ (the pitch of the periodic cladding) is assumed to be constant along the fiber and estimated to about 0.48 from scanning electron microscope images. The diameter variations of the fibers are proportional to those of the pitch, with $\Lambda_{min}=3.445~ \mu m$ corresponding to the minimum value of the diameter ($137 \mu m$, green line in Fig.~\ref{fig:fibres exp}(a)) and $\Lambda_{max}^{A,B,C}\simeq 4.48,~ 5.17$ and $6.03~ \mu m$, blue, black and red curves respectively, for the maximum values. As an example, the Group Velocity Dispersion (GVD) curve corresponding to the minimum pitch value has been calculated from Ref.\cite{saitho} and is represented in Fig.~\ref{fig:fibres exp}(b) as a  green curve. Its Zero Dispersion Wavelength (ZDW) is located at 1055 nm while those of the GVD curves corresponding to the maximum values of the diameters of fibers A, B and C are red-shifted to 1110 nm, 1136 nm and 1168 nm respectively (Fig.~\ref{fig:fibres exp}(b)). In order to give another illustration of the large dispersion variations induced in these fibers by varying their diameters, the maximum GVD values for fibers A, B and C have been calculated at a fixed wavelength (1052.5 nm) and compared to the background value. As can be seen in Fig. \ref{fig:fibres exp}(b), an increase of the diameters by a factor of only 1.27, 1.45 and 1.75, leads to a one order of magnitude improvement on the GVD values: 21, 29 and 35 respectively. Under such large variation of the fiber diameter, the GVD can no longer be considered as proportional to the pitch value, as was the case in Ref.\cite{droques13OL} for instance. This is illustrated in Fig.\ref{fig:fibres exp}(c), where the evolution of the GVD calculated at 1052.5 nm as a function of the fiber diameter is represented. It can be seen to be well approximated by an affine function in the range between 140$\mu$m and 180 $\mu$m, but not beyond. As a consequence, the shape of the GVD variations will be slightly different from the one of the diameter, specifically for fiber C. However, we checked numerically that this can be considered as relatively weak distortions that do not significantly impact the gain of the MI process. We can still consider that the key parameters remain the different heights in GVD of the Gaussian-like pulses in fibers A, B and C. In order to get a more complete picture of the impact of the fiber diameter variations on its guiding properties, the variation of the nonlinear coefficient is plotted as a function of the fiber diameter in Fig.~\ref{fig:fibres exp}(d). The most important feature to note here is that the amplitude of variation is much smaller, and only  the same order as the one of the diameter itself. Hence, these variations are more than one order of magnitude lower than those of the GVD and we have checked numerically that their impact on the MI process is negligible. Consequently, we can infer that these fibers represent a good prototype to validate our theoretical investigation in the previous sections, where only longitudinal GVD variations have been taken into account.
\begin{figure}
\begin{center}
\includegraphics[width=8cm]{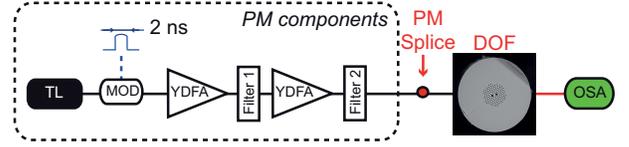}
\caption{Scheme of the experimental setup. MOD: intensity modulator; YDFA: ytterbium doped fiber amplifier; DOF: dispersion oscillating fiber; PM: polarization maintaining; TL: tunable laser; OSA: optical spectrum analyzer. \label{fig:experimental setup}}
\end{center}
\end{figure}
The experimental setup is schematized in Fig. \ref{fig:experimental setup}. The pump system is made of a continuous-wave tunable laser (TL) diode that is sent into an intensity modulator (MOD) in order to shape 2 ns square pulses at 1 MHz repetition rate. They are amplified by two Ytterbium-doped fiber amplifiers (YDFAs) at the output of which two successive tunable filters are inserted to remove the amplified spontaneous emission in excess around the pump. These quasi-CW laser pulses have been launched along the birefringent axis of the fibers. The pump peak power has been fixed to 6.5 W and the pump wavelength at 1052.5 nm for fiber A. The output spectrum recorded at its output is represented as a blue curve in Fig. \ref{fig:exp et simul} (a). Two MI side lobes, located at +/-4.8 THz appear on both sides of the pump. These experimental results have been compared with numerical simulations performed by integrating the generalized nonlinear Schr\"odinger equation. We used the GVD, $\beta_4$ and $\gamma$ variations calculated from the measured diameter values (see Figs. \ref{fig:fibres exp}). Other parameters are extracted from experiments and are listed in the caption of Fig. \ref{fig:exp et simul}. Note that we checked that except the longitudinal GVD variations, all other parameters can be assumed to be constant and equal to the average values. As can be seen in the blue curve in Fig. \ref{fig:exp et simul}(b), two symmetric MI side lobes also appear in the simulated spectrum, in a very good agreement with experiments. Their positions have been compared with the predictions of Equation~\ref{omega_l}, represented by green dashed lines in Fig. \ref{fig:exp et simul}(b) (calculated with $\beta_2^{average}=+0.59~ ps^2/km$). An excellent agreement is also obtained. In order to show that the MI gain is larger when the weight of the Dirac delta function is increased, we performed similar experiments in fibers B and C where the areas of the Gaussian pulses are larger than in fiber A. However, in experiments, due to the fact that the Dirac comb has been approximated with a series of Gaussian functions, changing their amplitudes also modifies the average value of the dispersion. As a consequence, MI side lobes would be generated at different frequency shifts. In order to keep constant the position of the MI side lobes, and then provide a correct comparison with the theoretical study, one have to take care to keep the average GVD value constant in all the fibers. To do so experimentally in fibers B and C, we slightly tuned the pump wavelength until the first MI side lobe is generated at 4.8 THz as in fiber A. As can be seen in Fig. \ref{fig:exp et simul}(a) (red and black curves), the position of the first MI side lobe in fibers B and C is indeed located at that frequency by tuning the pump wavelength to 1061.8 nm and 1067 nm, respectively. We can therefore consider that the average GVD values are very similar in the three fibers and hence that only the areas of the Gaussian functions, i.e. the equivalent of the Dirac weights, vary. The amplitudes of the first MI side lobes generated in fibers B and C are indeed larger compared to fiber A, as predicted by the theory. This is in pretty good agreement with numerical simulations (Fig.\ref{fig:exp et simul}(b)), where the same procedure was used. We found that the average values in fibers B and C are $\beta_2^{average}=+0.58~ ps^2/km$ and $\beta_2^{average}=+0.51~ ps^2/km$ respectively. The small discrepancy between these values is attributed to spurious longitudinal fluctuations arising during the drawing process. Indeed, as can be seen in Fig. \ref{fig:fibres exp}(a), the background over which Gaussian pulses are superimposed is not perfectly flat, and in fibre C, it is not horizontal. To counterbalance these imperfections, it was necessary to adjust the average dispersion values. We checked that with a series of perfect Gaussian pulses superimposed on a flat and horizontal background, the same average GVD value would be obtained. Furthermore, we can note that in fibers B and C, additional MI side lobes are generated due to the periodic modulation of the GVD (labelled MIi in \ref{fig:exp et simul}(a), up to 5 in fiber B). Their positions are also well predicted by numerical simulations and by Equation~\ref{omega_l} (green lines in \ref{fig:exp et simul}(b)). This excellent agreement confirms that their positions indeed scale approximately  as $\sqrt{l}$, $l$ being the side lobe order, that is the typical signature of the MI process occurring in dispersion oscillating fibers. It was already reported experimentally in Refs.\cite{droques13OL,copie2015}, with a sinusoidal variation of the GVD, modulated in amplitude or not and it is now illustrated in this paper with a Dirac delta comb. Moreover, in fibers B and C two symmetric side lobes that are not predicted by the theory appear around the pump at about 2.15 THz (labelled spurious side lobes in Fig.\ref{fig:exp et simul}(a)). They result from a non-phase matched four-wave mixing process involving the pump, the first and second MI side lobes. The energy conservation relation involving these waves predicts a frequency shift of 2.2 THz from the pump for the fourth wave, that is in good agreement with the shift of 2.15 THz measured experimentally.

\section{Conclusions}\label{s:conclusions}
Modulation instability has been investigated theoretically and experimentally in dispersion kicked optical fibers. An analytical expression of the parametric gain has been obtained allowing to predict the behavior of the MI process in such fibers. Specifically, it was shown that increasing the weights of the Dirac functions leads to larger MI gains for the first MI side lobe. We exploit the fact that the Dirac delta comb can be well approximated by a series of short Gaussian pulses in order to perform an experimental investigation using microstructured optical fibers. We then experimentally report, for the first time to our knowledge, multiple MI side lobes at the output of these dispersion kicked optical fibers. We demonstrate that they originate from the periodic variations of the dispersion. We also validate experimentally that increasing the height of the modulation leads to a larger gain for the first MI side lobe. This illustrates that optical fibers constitute an interesting platform to realize experimental investigations of fundamental physical phenomena.  

\section*{Acknowledgments}
The present research was supported  by the Agence 
Nationale de la Recherche in the framework of the Labex CEMPI (ANR-11-LABX-0007-01), Equipex FLUX (ANR-11-EQPX-0017), and by the projects TOPWAVE (ANR-13-JS04-0004), FOPAFE (ANR-12-JS09-0005) and NoAWE (ANR-14-ACHN-0014).

\end{document}